\documentclass[11pt,twoside]{article}
\usepackage{amsmath,amssymb}
\usepackage{psfig}
\usepackage{epsf}
\usepackage[english]{babel}
\usepackage{myletter}

\newcommand{\id}{\ensuremath{\text{d}}}
\newcommand{\ie}{{\itshape i.e.}}
\newcommand{\eg}{e.g.}
\newcommand{\dzero}{D\O}
\newcommand{\TeV}{\ensuremath {\text{\ Te\kern -0.1em V}}}
\newcommand{\GeV}{\ensuremath {\text{\ Ge\kern -0.1em V}}}
\newcommand{\invpb}{\ensuremath{\ \text{pb}^{-1}}}
\newcommand{\invfb}{\ensuremath{\ \text{fb}^{-1}}}
\newcommand{\pb}{\ensuremath{\ \text{pb}}}
\newcommand{\ra}{\ensuremath{\rightarrow}}
\newcommand{\pt}{\ensuremath{p_{\text{T}}}}
\newcommand{\ET}{\ensuremath{E_{\text{T}}}}
\newcommand{\MET}{\mbox{\ensuremath{\,\slash\kern-.7em E_{\text{T}}}}}
\newcommand{\Zob}{\ensuremath{\text{Z}}}
\newcommand{\Wb}[1]{\ensuremath{\text{W}^{#1}}}
\newcommand{\ppbar}{\ensuremath{\text{p}\bar{\text{p}}}}

\def\Journal#1#2#3#4{{\it #1} {\bf #2}, #3 (#4)}

\begin{document}

\begin{flushright}
{\large FERMILAB-CONF-06-318-E}
\end{flushright} 
\vskip 2em
\begin{center}
{\LARGE \Wb{} and \Zob{} Production at the Tevatron}
\vskip 1.5em
{\large
Michiel~P.~Sanders 
(for the \dzero{} and CDF Collaborations)\footnote{
Presented at the XXXIII International Conference on High Energy Physics, 
Moscow, Russian Federation, 26th July -- 2nd August 2006.}\\[.5em]}
{\itshape
Department of Physics, University of Maryland, College Park, MD 20742, USA}
\vskip 1em
\vskip 1.5em
\end{center}

\begin{abstract}
In this paper, recent experimental results on \Wb{} and
\Zob{} boson production at the Tevatron are described. These results
not only 
provide tests of the standard model, but are also sensitive to proton parton
distribution functions. 
\end{abstract}

\section{Introduction}

The Fermilab Tevatron
is a \ppbar{} collider, operating at a
centre-of-mass energy $\sqrt{s} = 1.96\TeV$. During Run~II (in Run~I, the
Tevatron operated at $\sqrt{s} = 1.8\TeV$), the two large, general
purpose experiments CDF and \dzero{} have collected an integrated
luminosity of approximately $1.5\invfb$ per experiment, at the time of writing. 
The results presented here have been obtained with integrated
luminosities of up to $350\invpb{}$. 

In the following, results on inclusive  \Wb{} and \Zob{} boson production
cross sections, the \Wb{} boson charge asymmetry and the
\Zob{} boson
rapidity distribution will be given. The experimental results will be
compared with predictions from the standard model, in particular
Quantum Chromodynamics (QCD), at 
next-to-leading-order (NLO) and at
next-to-next-to-leading-order (NNLO).

\section{\Wb{} and \Zob{} Boson Production}

In \ppbar{} collisions at the Tevatron energy, \Wb{} and \Zob{} bosons
are predominantly produced through the annihilation of a valence quark from
the proton and a valence anti-quark from the anti-proton. The probability for
such a $\text{q}\bar{\text{q}}$ annihilation to happen (\ie{}, the cross
section) is 
perturbatively calculable from the standard model. Corrections due to
gluon exchange in the $\text{q}\bar{\text{q}}$ pair or gluon
emission from the 
(anti-) quark are known from QCD.
The probability to find a quark of a certain momentum inside
the proton is known from experimentally determined parton distribution
functions (PDF's). 
A measurement of inclusive or differential 
\Wb{} and \Zob{} boson production cross sections
therefore provides tests of both QCD and PDF's. 

In addition, many other physics processes of interest, such as
top-quark decays, involve electroweak bosons. A precise
experimental understanding of these bosons is thus a benchmark for
other analyses. 
Also, since theoretical predictions for \Wb{} and \Zob{} boson production
are quite accurate, a measurement of the production rate can serve as
a tool to determine the integrated luminosity. This will be especially
useful at the Large Hadron Collider. 

\subsection{Experimental Signatures and Backgrounds}

The electroweak bosons are detected in their leptonic decay
modes\footnote{Hadronic decays are very hard to
  select, due to the large multi-jet background from QCD processes.}.
In case of the \Wb{} boson, the experimental signature is events with a single
high \pt{} lepton (electron or muon) and a large missing transverse
energy (\MET{}) due to the undetected neutrino. 
With cuts on \pt{} and \MET{} typically set at 20 to 25\GeV{},
about 1~million candidates are selected
per inverse femtobarn of integrated luminosity.
Background in this sample is small, and is composed of
so-called QCD background\footnote{``QCD'' here refers to the
  production of jets in the final state. Of course, the initial state
  in electroweak boson production is also governed by QCD.} and
electroweak boson background. QCD background in the muon channel is dominated
by semi-leptonic decays of charm and bottom hadrons, and amounts to
approximately 1\% of the \Wb{} boson candidate sample. In the electron
channel, jets with an electron signature form the QCD background
($\sim 1\%$). The electroweak boson backgrounds are due to leptonic
decays of \Zob{} bosons where one of the leptons is not identified
($\sim 5\%$ in the muon channel, $\sim 0.5\%$ in the electron
channel), and $\Wb{}\ra\tau\nu$ decays ($\sim 2\%$ in both channels).

\Zob{} boson decays are selected as events with two high \pt{} leptons 
($\pt > 15 - 25\GeV$). For each inverse femtobarn of integrated
luminosity, this gives a sample of about $10^5$ candidates.
Backgrounds are very small. QCD
background is $\sim 0.5\%$ in the muon channel and $\sim 1\%$ in the
electron channel. The only other significant source of background
is from $\Zob \ra \tau\tau$ decays ($< 0.5\%$).

\subsection{Cross Section Measurements}

The \Wb{} or \Zob{} boson
production cross section (times the branching fraction) can
be determined as follows:  
\begin{equation}
\sigma(\ppbar \ra \Wb{}/\Zob{})\times B( \Wb{}\ra \text{l}\nu / \Zob{}\ra \text{ll})
             =
        \frac{N_{\text{candidates}} - N_{\text{background}}}
             {\epsilon A  \int {\cal L} \id t},
\label{eq:crosssection}
\end{equation}
where the number of candidates is given by
$N_{\text{candidates}}$. The backgrounds (given by
$N_{\text{background}}$), as described in the previous section, are
estimated from the data itself (for the QCD backgrounds) or from
Monte Carlo simulations of standard model predictions (for the
electroweak boson backgrounds). The efficiency $\epsilon$ to correctly
identify and select a \Wb{} or \Zob{} boson is
typically determined from the data, with an uncertainty of $1-2\%$.  
The experimental acceptance $A$ is calculated from Monte
Carlo simulations of \Wb{} or \Zob{} boson production. The uncertainty on
the acceptance is $\sim 1.5\%$, and is dominated by
uncertainties in the PDF's. Finally, the integrated luminosity ($\int
{\cal L} \id t$) is measured by independent luminosity systems.

\begin{figure}[htb]
\centerline{\psfig{file=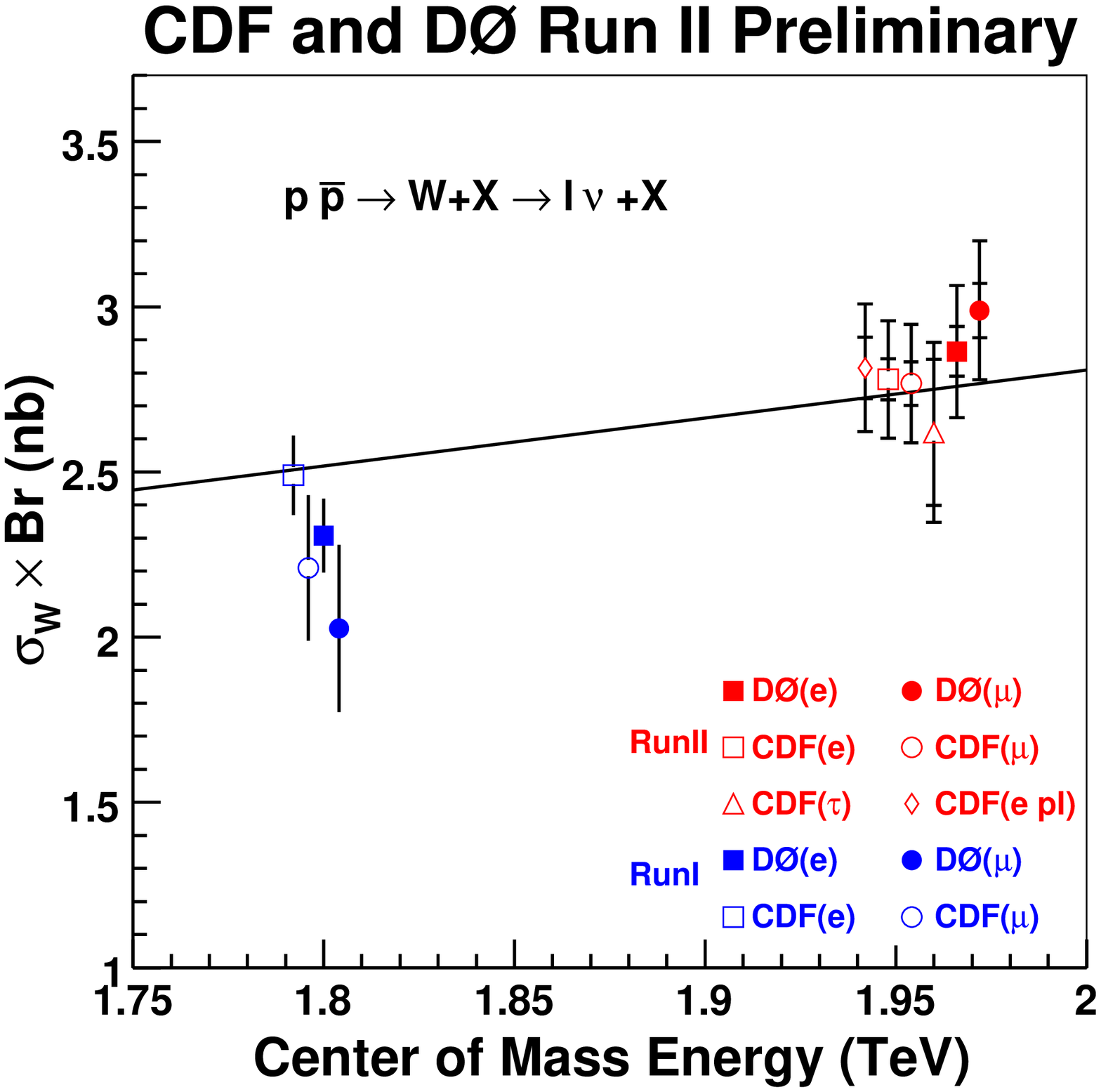,width=.5\textwidth}}
\centerline{\psfig{file=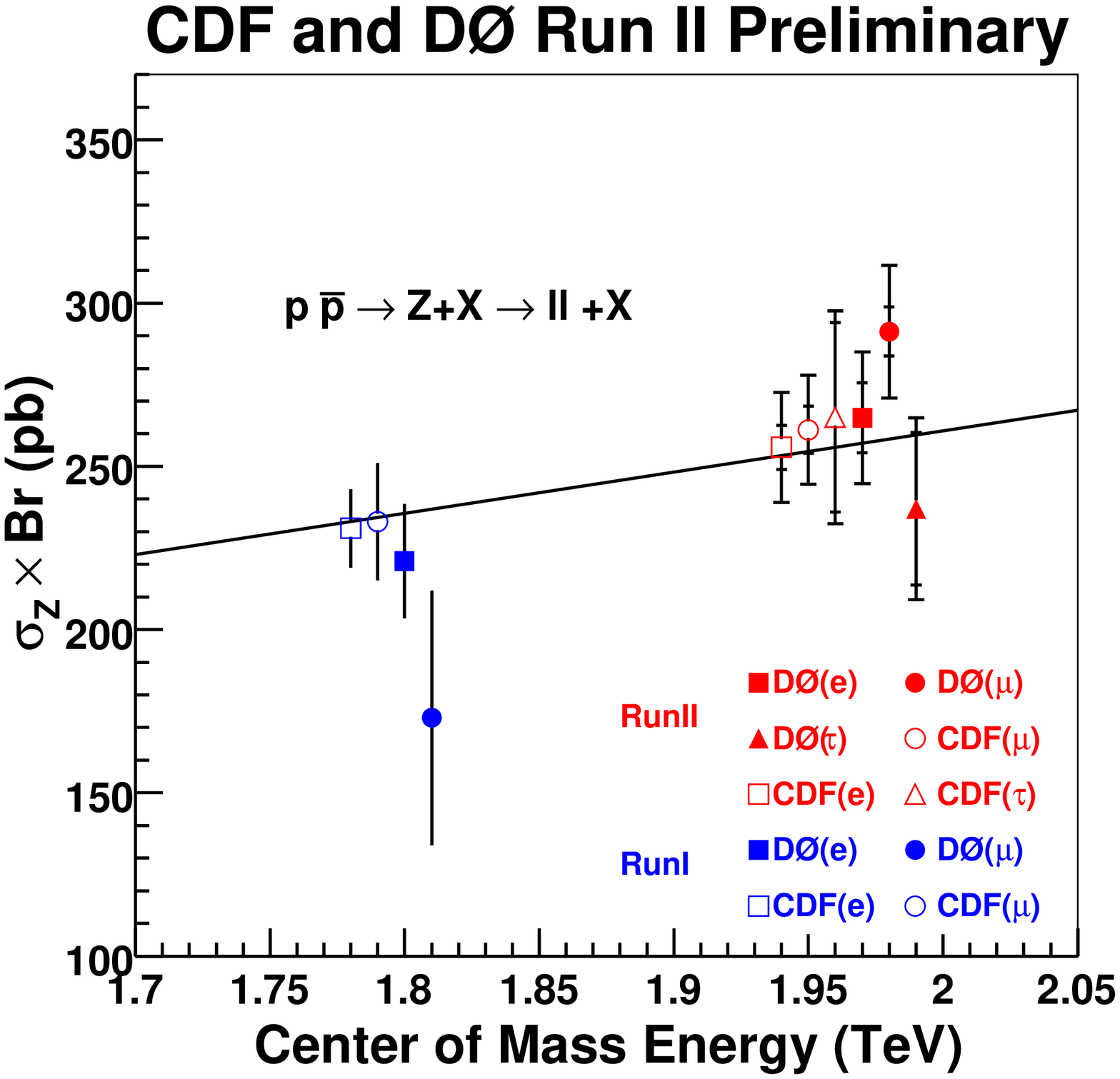,width=.5\textwidth}}
\caption{Measured and predicted cross section times branching fraction
  for \Wb{} (top)
  and \Zob{} (bottom) boson production, as a function of centre-of-mass
  energy.}
\label{fig:WZcs_vs_sqrts}
\end{figure}

Current results on \Wb{} and \Zob{} boson cross section measurements at the
Tevatron are shown in Figure~\ref{fig:WZcs_vs_sqrts}.  The largest
systematic uncertainty on these measurements ($\sim 6\%$) is
due to the measurement of the integrated luminosity 
($\int {\cal L} \id t = 72 - 350\invpb{}$
for the Run~II points).  The solid line in these figures represents an NNLO
QCD prediction\cite{bib:CS_Hamberg_theory}. There is good agreement between the
experimental points and the prediction. Note that cross sections
in the $\tau$ decay channel have also been measured. These are
important benchmarks for Higgs and new physics searches.

\subsection{Cross Section with Forward Electrons}

Typically, PDF's are parameterized as a function of 
$x$, the fraction of the proton momentum carried by the quark, and
$Q^2$, the momentum transfer between the quark and anti-quark. 
At leading order, the rapidity of a \Wb{} boson ($y_{\Wb{}}$) is
related to $x$ via $x = \frac{m_{\Wb{}}}{\sqrt{s}}
e^{(\pm)y_{\Wb{}}}$. 
A measurement of the \Wb{} boson production cross section at large $y_{\Wb{}}$
is therefore sensitive to PDF's at small and large $x$. In its turn,
the pseudorapidity $\eta$ of the lepton from a \Wb{} boson decay is
correlated with the rapidity of the \Wb{}
boson. 
This implies that leptons at large $\eta$ (\ie{}, in the forward
detectors) come from \Wb{} bosons with a large rapidity on average, whereas
leptons at small $\eta$ (in the central detectors) are due to \Wb{} bosons
with small rapidity.

The CDF collaboration has exploited this in a measurement of the \Wb{}
boson production cross section using electrons in the forward
calorimeters ($1.2 < |\eta| < 2.8$).
After a full correction for the limited acceptance of the forward
calorimeters, the resulting cross section times branching fraction is:
\begin{equation}
\sigma(\ppbar \ra \Wb{}) \times B(\Wb{} \ra \text{e}\nu) = 2796
        \pm 13 (\text{stat}) 
        ^{+95}_{-90} (\text{sys}) 
        \pm 162 (\text{lum})\pb.
\end{equation}
This number is in good agreement with the cross section from electrons
in the central detector ($|\eta| < 1.0$) and with the predicted cross
section. Note that this measurement is also shown in
Figure~\ref{fig:WZcs_vs_sqrts} (``CDF (e pl)''). 

A first-order test of PDF's is obtained by taking the ratio of the
visible cross section (defined as $\sigma_{\text{vis}} =
\sigma_{\Wb{}} \times A$) in the central and forward
detectors. The cross section uncertainty due to the PDF's cancels
in this ratio, and the uncertainty due to the integrated
luminosity is reduced. This cross section ratio is equivalent to the
ratio of \Wb{} boson acceptances for forward and central electrons.
The CDF result is:
\begin{eqnarray}
\sigma_{\text{vis}}^{\text{cent}} / \sigma_{\text{vis}}^{\text{forw}} 
                                  &=& 0.925 \pm 0.033 ,
\label{eq:cs_ratio_measured} \\
A^{\text{cent}} / A^{\text{forw}} &=& 0.924^{+0.023}_{-0.030} 
   \quad (\text{CTEQ}), 
\label{eq:acc_ratio_CTEQ} \\
A^{\text{cent}} / A^{\text{forw}} &=& 0.941^{+0.011}_{-0.015} 
   \quad (\text{MRST}).
\label{eq:acc_ratio_MRST}
\end{eqnarray}
The uncertainties on the expected ratios are dominated by the PDF
uncertainty.
It can be seen that the measured ratio
(Eq.~\ref{eq:cs_ratio_measured}) is in excellent agreement with the
predictions using the CTEQ PDF's (Eq.~\ref{eq:acc_ratio_CTEQ}) and in
good agreement with the MRST
prediction (Eq.~\ref{eq:acc_ratio_MRST}).
It is clear that just this $\sigma_{\Wb{}}(\eta_{\,\text{l}})$
measurement is already sensitive to the PDF's.

\section{\Wb{} Boson Charge Asymmetry}

On average, the u-quark in the proton carries a larger fraction of the
proton's momentum than the d-quark. This implies that at production, 
a $\Wb{+(-)}$ boson is typically boosted in the $\text{p}(\bar{\text{p}})$
direction. This leads to a charge asymmetry, as a function of $y_{\Wb{}}$.
A measurement of this asymmetry is directly sensitive to the PDF's. 

The momentum of the neutrino in a \Wb{} boson decay is not fully
reconstructible, which implies that $y_{\Wb{}}$ is not fully
reconstructible. However, some of the original asymmetry is preserved
in the pseudorapidity distribution of the decay-lepton.

Both the \dzero{} (muon channel) and CDF (electron channel) 
collaborations have measured the lepton
charge asymmetry, defined as:
\begin{equation} 
A(y) = \frac{N^+(y) - N^-(y)}{N^+(y) + N^-(y)}.
\end{equation}
These measurements probe a region in $(x,Q^2)$ that is not probed by,
\eg{}, HERA experiments. The main experimental challenge is to keep
the lepton-charge misidentification rate low ($\sim 0.01\%$ for \dzero{}'s
muon channel; $\sim 1\%$ for CDF's electron channel), and to
understand possible charge-dependencies of selection efficiencies. 

\begin{figure}[hbt]
\centerline{\psfig{file=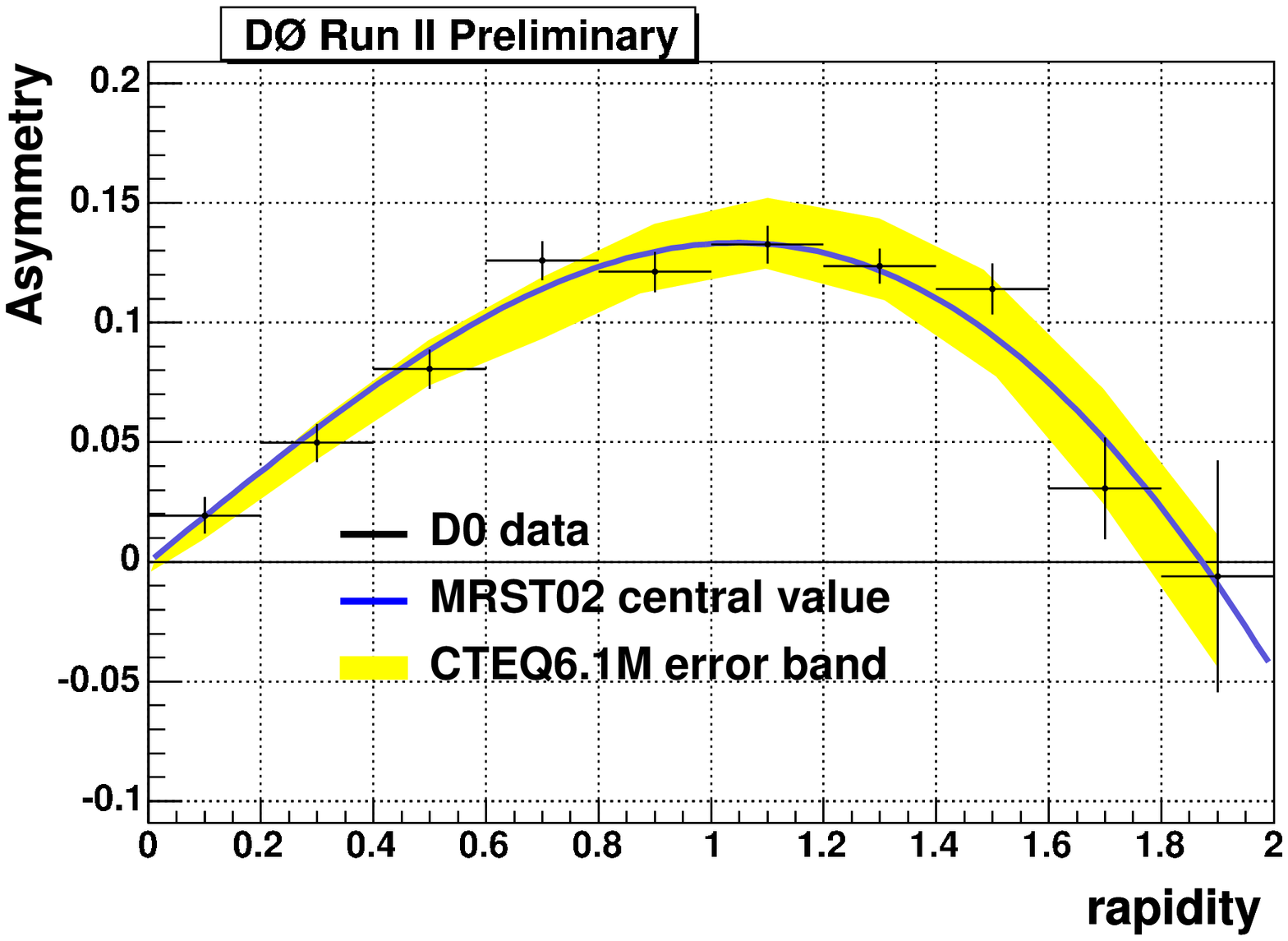,width=.65\textwidth}}
\caption{Measured and predicted lepton charge asymmetry, as a function
  of lepton rapidity.
  The theoretical curves are based on NLO calculations (\textsc{resbos}).}
\label{fig:Wasym_D0_muon}
\end{figure}

The result of \dzero{}'s measurement, based on $\int {\cal L} \id t
\simeq 230\invpb{}$, is given in Figure~\ref{fig:Wasym_D0_muon}.
Even though the uncertainties on the experimental points are dominated by the
finite data-sample-size, some sensitivity to PDF's is already obtained
(assuming
the \Wb{} boson decays according to the standard model). 
Thus, with more data, this measurement will constrain the
PDF's.

A description of the CDF measurement ($\int {\cal L} \id t \simeq 170
\invpb{}$) can be found in Ref.~\cite{bib:Wasym_CDF_electron}. CDF
split up the final data sample in two bins ($25 < \ET(\text{e}) < 35
\GeV$ and $35 < \ET(\text{e}) < 45
\GeV$) to get additional PDF sensitivity. 

\section{\Zob{} Boson Production Rapidity}

\begin{figure}[hbt]
\centerline{\psfig{file=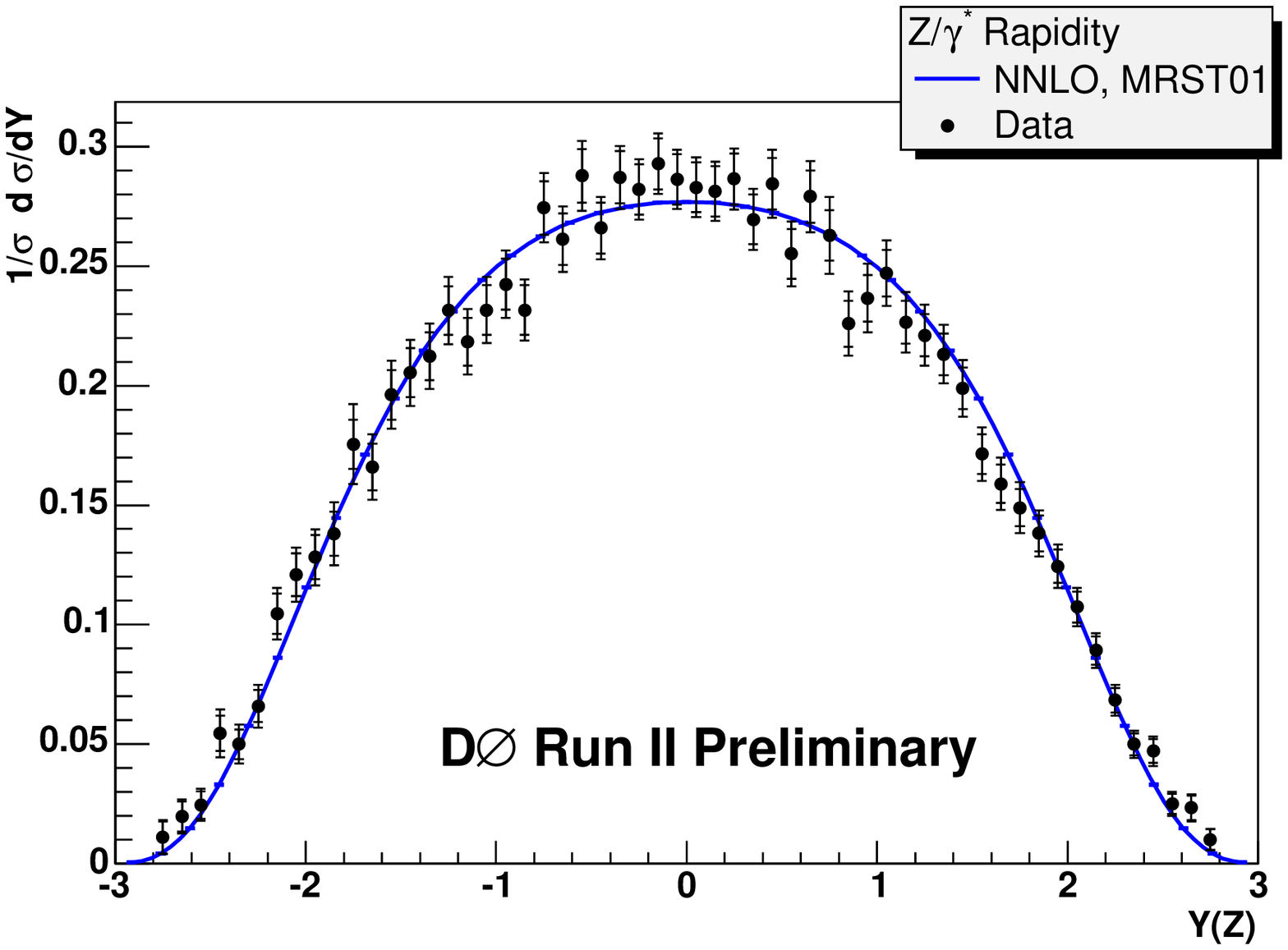,width=.65\textwidth}}
\caption{Measured and predicted \Zob{} boson rapidity distribution.}
\label{fig:Zrapidity_D0}
\end{figure}

The rapidity of a \Zob{} boson ($y_{\Zob{}}$) produced in a \ppbar{} collision is
also sensitive to the u- and d-quark momentum fractions in the
proton. As opposed to the \Wb{} boson rapidity, the rapidity of the
\Zob{} boson is fully
reconstructible. However, u- and d-quark contributions cannot be
separated. The result of a measurement by the \dzero{}
collaboration of $y_{\Zob{}}$ in the electron
channel ($\int {\cal L} \id t
\simeq 340\invpb{}$) is given in Figure~\ref{fig:Zrapidity_D0}. A good
agreement with an NNLO prediction\cite{bib:Zrapidity_anastasiou} is
observed, but the data sample is still too small to be sensitive to
PDF's. 

\section{Conclusion and Outlook}

Precision measurements of inclusive and differential \Wb{} and \Zob{} boson
production cross sections at the Tevatron have been performed. The
results agree well with standard model predictions. Measurements of
lepton charge asymmetries as a function of lepton rapidity are
already sensitive to PDF's. In the near
future, an order of magnitude more data will be analyzed. This will
provide a better determination of PDF's, which in its turn is
important for reducing systematic uncertainties on, for example, the \Wb{}
boson mass measurement.


\end{document}